\documentclass[draftclsnofoot]{IEEEtran}
\usepackage{graphicx}
\usepackage{epstopdf}
\usepackage{subfigure}
\usepackage{booktabs}
\usepackage{caption}
\usepackage{color}
\usepackage{bm}
\usepackage{CJK}
\usepackage{indentfirst}
\usepackage{amsmath}
\usepackage{amssymb}
\usepackage{float}
\usepackage{multirow}
\usepackage{algorithm}
\usepackage{algpseudocode}
\usepackage{amsmath}
\usepackage{flushend}
\usepackage{cases}
\usepackage{geometry}
\newtheorem{proposition}{Proposition}
\newtheorem{corollary}{Corollary}

\begin{document}

\title{Rotatable Block-Controlled RIS: Bridging the Performance Gap to Element-Controlled Systems}

\author{\IEEEauthorblockN{Weicong~Chen, Xinyi~Yang, Chao-Kai~Wen, Wankai~Tang, Jinghe~Wang, Yifei Yuan, Xiao~Li, and~Shi~Jin}\\
\thanks{{Weicong Chen, Xinyi Yang, Wankai Tang, Jinghe Wang, Xiao Li and S. Jin are with the National Mobile Communications Research Laboratory, Southeast University,
Nanjing, 210096, P. R. China (e-mail: cwc@seu.edu.cn; yangxinyi@seu.edu.cn; tangwk@seu.edu.cn; wangjh@seu.edu.cn; li\_xiao@seu.edu.cn; jinshi@seu.edu.cn).} }
\thanks{Chao-Kai Wen is with the Institute of Communications Engineering, National Sun Yat-sen University, Kaohsiung 80424, Taiwan. (e-mail: chaokai.wen@mail.nsysu.edu.tw).}
\thanks{Yifei Yuan are with the China Mobile Research Institute, Beijing 100053, China (e-mail: yuanyifei@china mobile.com).}
}

\maketitle
\begin{abstract}

The passive reconfigurable intelligent surface (RIS) requires numerous elements to achieve adequate array gain, which linearly increases power consumption (PC) with the number of reflection phases. To address this, this letter introduces a rotatable block-controlled RIS (BC-RIS) that preserves spectral efficiency (SE) while reducing power costs. Unlike the element-controlled RIS (EC-RIS), which necessitates independent phase control for each element, the BC-RIS uses a single phase control circuit for each block, substantially lowering power requirements. In the maximum ratio transmission, by customizing specular reflection channels through the rotation of blocks and coherently superimposing signals with optimized reflection phase of blocks, the BC-RIS achieves the same averaged SE as the EC-RIS. To counteract the added power demands from rotation, influenced by block size, we have developed a segmentation scheme to minimize overall PC. Furthermore, constraints for rotation power-related parameters have been established to enhance the energy efficiency of the BC-RIS compared to the EC-RIS. Numerical results confirm that this approach significantly improves energy efficiency while maintaining performance.  
\end{abstract}

\section{Introduction}

Research focus in wireless communication is shifting towards the next generation to meet the ambitious 6G vision of ultra-high data rate services. Key technologies such as ultra-massive multiple-input multiple-output (MIMO) and terahertz communications, evolving from 5G's massive MIMO and millimeter wave (mmWave) communications \cite{6G}, face escalating hardware costs and power consumption (PC) challenges. The low-cost, low-power reconfigurable intelligent surface (RIS) is emerging as a vital technology for enhancing capacity in 6G.


RIS applications in wireless communications have already gained significant attention \cite{E. Basar}. Studies show that the RIS-assisted symbol level precoding improves the energy efficiency (EE) for over $40\%$ when compared to the zero forcing (ZF) precoding scheme \cite{App_EE}. An RIS-assisted cell-free massive MIMO system can achieve higher performance by jointly optimizing the user's power control and RIS's passive beamforming \cite{Q. Wang-IRS}. Furthermore, integrating RIS into satellite-terrestrial relay networks can cut total transmit power by jointly optimizing the beamforming vectors and phase shifters \cite{app-22-ZLin}. In symbiotic cognitive radio networks, a resource allocation framework was proposed to maximize the sum rate of primary and secondary networks using RIS \cite{TVT-2}. Moving beyond the need for instantaneous channel state information (CSI), one study used statistical CSI to maximize averaged spectral efficiency (SE), addressing practical channel estimation challenges in RIS-assisted systems \cite{Y. Han-LIS}. On the basis of statistical CSI, RIS's unique ability to modify transmission channels was highlighted in \cite{CC-1} and subsequent works, enhancing multiplexing in mmWave communications, reducing CSI feedback overhead, and adapting various transmission schemes.

Despite the enormous application potentials of RIS in wireless communications, the issue of PC increasing linearly with the number of RIS elements cannot be ignored when distributed massive RISs are deployed. Recently, fluid antennas \cite{FA-kit} and movable antennas \cite{MA}, which can constitute a virtual MIMO using fewer flexible-position antennas, were introduced to wireless communications. Moving a single antenna by one position may consume less power than adding an additional antenna \cite{FP-MIMO}. Considering that the wireless channel is determined by the position of antennas, these position-flexible antenna techniques can physically reshape the channel, making a departure from channel customization schemes \cite{CC-1} realized by manipulating the propagation of electromagnetic waves.

Leveraging the physical channel reshaping capability facilitated by flexible positioning, this study conceives a novel RIS architecture, namely, \emph{rotatable} block-controlled RIS (BC-RIS)\footnote{A block of $M$ elements in the BC-RIS is connected to a single phase control circuit that generates a common reflection phase across the block. Compared to the EC-RIS that uses individual phase control circuits to assign diverse reflection phases to each element, the BC-RIS consumes $1/M$ of the power for the phase control circuit \cite{B. Zheng}. However, when optimizing phases to achieve the maximal SE, it loses $M-1$ degrees of freedom for reflection phase design in each block, resulting in a potential SE loss \cite{BC-RIS}.}, to approach the SE performance of the traditional element-controlled RIS (EC-RIS)  \cite{Y. Han-LIS} with lower consumption. In the maximum ratio transmission (MRT), we propose to eliminate the SE gap of EC-RIS \cite{Y. Han-LIS} and BC-RIS \cite{BC-RIS} by designing the rotation angles of RIS blocks. The additional rotation PC related to the block size is considered and minimized by an RIS block segmentation scheme. To surpass the EE of the EC-RIS, we provide guidance on determining the rotation-related power coefficients for the hardware implementation of the BC-RIS. Numerical results verify that the BC-RIS can achieve the same SE as the EC-RIS with lower power, especially at higher frequency band.



\section{System Model}\label{sec:2}
\begin{figure}
	\centering
	\includegraphics[width=0.46\textwidth]{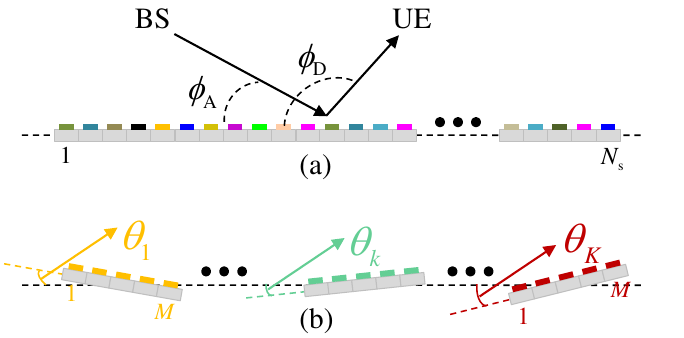}
	\caption{RIS with different control mechanisms: (a) EC-RIS with independently tunable reflection phases for each element; (b) BC-RIS with rotatable blocks constituted by elements sharing the same reflection phase. The angle increases clockwise, and the direction from the left end of the array is defined as zero.}
	\label{Fig.system-model} 
	\vspace{-0.5cm}
\end{figure}

This study considers a downlink MISO system where a BS with $N_{\rm b}$ antennas transmits signals to a single-antenna user equipment (UE) via a reflection link cascaded by the RIS. Focusing on the benefit of BC-RIS, we assume that both the BS and the RIS employ uniform linear arrays (ULAs) with half-wavelength spacing for simplification, and we leave the uniform planar array cases for future work.  Given that the EC-RIS exhibits PC that increases linearly with the number of elements, the BC-RIS is employed to achieve potentially higher EE in this study, as illustrated in Fig. \ref{Fig.system-model}. We assume that the $N_{\rm s}$-element RIS is divided into $K$ rotatable block, with each block comprises $M$ elements that share a single control circuit for reflection phase configuration, i.e., ${N_{\rm s} = K \times M}$.

In the downlink, the received signal at the UE is given by
\begin{equation}\label{Eq-1}
	y = {{\bf{h}}^H}{\bf{f}}x + z,
\end{equation}
where ${\bf{h}} \in {{\mathbb C}^{{N_{\rm{b}}} \times 1}}$ represents the channel between the BS and the UE; ${\bf{f}}\in {{\mathbb C}^{{N_{\rm{b}}} \times 1}}$ and $x \in {\mathbb C}$ are the precoding vector and the transmit signal at the BS, respectively; $z \sim {\mathcal{CN}}(0,{\sigma}^2)$ denotes the additive noise with power $\sigma^2$. The power constraint at the BS is defined as $P$.

\subsection{Channel Model}

In scenarios where the direct link between the BS and the UE is severely obstructed, introducing the RIS can provide an artificial transmission link. In this context, the channel between the BS and UE can be expressed as
\begin{equation}\label{Eq:h}
	{{\bf{h}}^H} = {{\bf{g}}^T}{\rm{diag}}\left( {\bm{\gamma }} \right){\bf{G}},
\end{equation}
where ${\bf{g}} \in {{\mathbb C}^{{N_{\rm{s}}} \times 1}}$ and ${\bf{G}} \in {{\mathbb C}^{{N_{\rm{s}}} \times {N_{\rm{b}}}}}$ denote the RIS--UE channel and the BS--RIS channel, respectively. With the BC--RIS, the reflection vector ${\bm{\gamma }}\in {{\mathbb C}^{{N_{\rm{s}}} \times 1}}$ is given by
\begin{equation}\label{Eq:gamma_v}
	{\bm{\gamma }} = {\left[ {{e^{j{\gamma _1}}},{e^{j{\gamma _2}}}, \ldots ,{e^{j{\gamma _K}}}} \right]^T} \otimes {{\bf{1}}_{M \times 1}},
\end{equation}
where $\gamma_k$ represents the common reflection phase of elements in the $k$-th block, and ${{\bf{1}}_{M \times 1}}$ is a vector of all ones.

When the RIS is located in the far-field of both the BS and the UE, the Rician channel model is adopted to describe ${\bf{g}}$ and ${\bf{G}}$. In this context, the BS-RIS channel is modeled as
\begin{equation}\label{Eq:G}
	{\bf{G}} = \sqrt {\frac{{{\kappa _1}}}{{{\kappa _1} + 1}}} {\bar{\bf G}} + \sqrt {\frac{1}{{{\kappa _1} + 1}}} {\tilde{\bf G}},
\end{equation}
where $\kappa_1$ is the Rician factor, ${\bar{\bf G}}$ and ${\tilde{\bf G}}$ are the line-of-sight (LoS) and non-LoS (NLoS) channel component, respectively. The LoS channel component ${\bar{\bf G}}$ can be expressed as ${\bar{\bf G}}={\sqrt{N_{\rm b}N_{\rm s}}} \  {\bf{a}}_{{\rm s}}( {\phi _{\rm{A}},{\bm{\theta }}} ){\bf{a}}_{\rm{b}}^H( {\varphi _{\rm{D}}} )$, where the normalized array response vector at the BS is given by
\begin{equation}\label{Eq:ARV-BS}
{{\bf{a}}_{\rm{b}}}\left( {{\varphi _{\rm{D}}}} \right) = \frac{1}{{\sqrt {{N_{\rm{b}}}} }}{\left[ {1,{e^{j\pi\cos {\left( {{\varphi _{\rm{D}}}} \right)}}}, \ldots ,{e^{j{\left( {{N_{\rm{b}}} - 1} \right)}\pi\cos {\left( {{\varphi _{\rm{D}}}} \right)}}}} \right]^T}
\end{equation}
with ${{\varphi _{\rm{D}}}}$ being the angle of departure (AoD) at the BS. The NLoS channel component ${\tilde{\bf G}}$ follows the Rayleigh fading with zero mean and unit variance.

For the RIS with rotatable blocks, the normalized array response vector is
\begin{equation}\label{Eq:ARV-RIS}
	{\bf{a}}_{{\rm s}}{\left( {\phi_{\rm A} ,{\bm{\theta }}} \right)}=\frac{1}{{\sqrt {{N_{\rm{s}}}} }}{\left[ {{\bf{a}}_{{\rm s}_1}^T{\left( {\phi_{\rm A},{\theta _1}} \right)}, \ldots ,{\bf{a}}_{{\rm s}_K}^T{\left( {\phi_{\rm A},{\theta _K}} \right)}} \right]^T},
\end{equation}
where ${\phi_{\rm A}}$ is the angle of arrival (AoA) at the RIS, ${\bm{\theta }}=[\theta_1,\theta_2,\ldots,\theta_K]$ is the rotation angle vector of $K$ RIS blocks, and ${\bf{a}}_{{\rm{s}}_k}( {\phi_{\rm A},{\theta _k}} )$ is the physically tunable array response vector of the $k$-th block with the $m$-th element given by
\begin{equation}\label{Eq:ARV-sub-RIS}
	{\big[ {{{\bf{a}}_{{{\rm{s}}_k}}}{\left( {{\phi _{\rm{A}}} , {\theta _k}} \right)}} \big]_m} = {e^{j\left({\Phi _k}+{\Theta _{k,m}}\right)}}
\end{equation}
where ${\Phi _k}={( {k - 1} )M\pi \cos ( \phi_{\rm A}  )}$  represents the global phase difference between the rotate center of the first and the $k$-th blocks, and ${\Theta _{k,m}} = ( {m - \frac{{M + 1}}{2}} )\pi \cos ( {{\phi _{\rm{A}}} - {\theta _k}} )$ denotes the local phase difference between the $m$-th element and the center point of the $k$-th block.

Similar to ${\bf{G}}$, the RIS--UE channel ${\bf{g}}$ is modeled as
\begin{equation}
	{\bf{g}} = \sqrt {\frac{{{\kappa _2}}}{{{\kappa _2} + 1}}} {\bar{\bf g}} + \sqrt {\frac{1}{{{\kappa _2} + 1}}} {\tilde{\bf g}}
\end{equation}
where $\kappa _2$ is the Rician factor, ${\bar{\bf g}} = \sqrt {{N_{{\rm s}}}} \ {\bf{a}}_{{\rm s}}( {{\phi _{\rm{D}}},{\bm{\theta }}} )$ is the LoS component with AoD being ${\phi _{\rm{D}}}$, and ${\tilde{\bf g}}\sim {\mathcal{CN}}(0,{\bf I})$ is the NLoS component. The array response vectors at the RIS depend on the rotation angles of the RIS blocks, which means that ${\bf{G}}$ and ${\bf{g}}$ are customizable.

\subsection{Power Consumption Model}
In this study, the varactor-based RIS is adapted to realize continuously reflection phase control. Another advantage of varactor-based RIS is that each element consumes negligible energy \cite{J.Wang}. On the basis of the PC modeling and practical measurement results in \cite{J.Wang}, the total system PC with the varactor-based EC-RIS is given by
\begin{equation}
	P_{\rm EC} = P_0+\xi P  +  N_{\rm s}P_1 ,
\end{equation}
where $P$ is the transmit power of the BS, $\xi$ is the power amplifier efficiency, $P_{1}$ is the PC of a single ``reflection phase'' control circuit, and $P_0=P_{\rm BS}+P_{\rm UE}+P_{\rm RIS}$ denotes sum of the constant circuit PC at the BS, UE, and RIS. When the BC-RIS is adopted, the number of reflection phase control circuits can be reduced from $N_{\rm s}$ to $K$. Note that the PC of each phase control circuit remains unchanged, as the same voltage is applied in parallel across all elements, while the current passing through each element is virtually zero. However, the PC of the rotation mechanism must also be considered for the proposed rotatable BC-RIS. This PC is primarily composed of circuit control and mechanical rotation. We denote $P_2$ is the PC of a single ``rotate'' control circuit. To rotate a $M$-element block by $|\theta|$, the required mechanism PC can be calculated by 2${\sum_{m = 0}^{ {(M - 1)}/{2} } {m\left| {{\theta}} \right|{P_{{\rm{unit}}}}} }$, where ${P_{{\rm{unit}}}}$, determined by the weight and aperture of the block, denotes the PC for rotating an element one unit distance from the rotation center by one unit angle. Thus, the system PC with the BC-RIS can be modeled as
\begin{equation}
	P_{\rm BC} = P_0 +\xi P  +  K\left(P_1+P_2\right)+2 \sum_{k = 1}^K {\sum_{m = 0}^{ {(M - 1)}/{2} } {m\left| {{\theta _k}} \right|{P_{{\rm{unit}}}}} } ,
\end{equation}
Note that when $K=N_{\rm s}$, i.e., the block size $M = 1$, the rotation mechanism is unnecessary, and $P_2 = 0$.

\subsection{Performance Metrics}
When the MRT ${\bf f}= {\bf{h}}/{\left\| {\bf{h}} \right\|}$ is adopted and the transmission satisfies ${\mathbb E}\{ \left\| {{\bf{f}}x} \right\|^2  \}=P$, the downlink SE can be expressed as
\begin{equation}\label{Eq:SE}
	S_{\rm XC} = {\log _2}\left( {1 + {P{\left\|{{\bf{g}}^T}{\rm{diag}}\left( {\bm{\gamma }} \right){\bf{G}} \right\|_F^2}}/{{{\sigma ^2}}}} \right),
\end{equation}
where the subscript ${\rm XC}\in \{{\rm EC, \ BC}\}$ indicates whether the RIS is element-controllable or block-controllable. The energy efficiency is then given by
\begin{equation}
	E_{\rm XC} = {S_{\rm XC}}/{P_{\rm XC}}.
\end{equation}

Intuitively, when the EC-RIS is replaced with the block-controlled counterpart, the design degree of freedom for reflection phases is reduced from $N_{\rm s}$ to $N_{\rm s}/M$. Because the block size $M \ge 2$, the SE degrades with $M$ if channels ${\bf g}$ and ${\bf G}$ keep unchanged. Fortunately, the rotatable block provides potentials to modified ${\bf g}$ and ${\bf G}$, as can be seen in \eqref{Eq:ARV-RIS}. Proceed from this point, in the following we investigate how to approach and achieve the same performance of the EC-RIS by designing reflection phases and rotation angles of the BC-RIS to finetune the channel.


\section{Performance Analysis}\label{sec:3}
The availability of CSI is essential for designing the RIS. However, obtaining the instantaneous CSI of $\bf g$ and $\bf G$ is challenging due to the lack of baseband signal processing capability at the RIS. Therefore, we use statistical CSI, i.e., the AoA and AoD of the LoS paths, for the RIS design and accordingly evaluate the averaged SE and EE, denoted as ${\mathbb{E}}\{S_{\rm XC}\}$ and ${\mathbb{E}}\{E_{\rm XC}\}$, respectively.

\subsection{Averaged Spectral Efficiency Analysis}
To provide intuitive insights into how the design of the RIS affects the averaged SE, we establish an upper bound in the following proposition:

\begin{proposition}\label{Pro-1}
	When the MRT is adopted for rotatable BC-RIS-assisted MISO systems, the averaged SE is upper bounded by
	{{\begin{equation}\label{Eq:SE-u}
		{{{S_{\rm BC, u} = {\log _2}{\left( {1 + \frac{P}{{{\sigma ^2}}}{\left( {{C_1}{{\left| {\sum\limits_{k = 1}^K {{e^{j{R_{1,k}}}}\sum\limits_{i = 1}^M {{e^{ - j{R_{2,k,i}}}}} } } \right|}^2} + {C_2}} \right)}} \right)}},}}
	\end{equation}}}
where ${C_1} = \frac{{{N_{\rm{b}}}{\kappa _1}{\kappa _2}}}{{\left( {{\kappa _1} + 1} \right)\left( {{\kappa _2} + 1} \right)}}$ and $C_2 = \frac{{{N_{\rm{b}}}{N_{{\rm s}}}\left( {{\kappa _2} + {\kappa _1} + 1} \right)}}{{\left( {{\kappa _1} + 1} \right)\left( {{\kappa _2} + 1} \right)}}$ are constants. The reflection phase-determined $R_{1,k}$ is given by
\begin{equation}
	{R_{1,k}} = {\gamma _k}+\left( {k - 1} \right)M\pi \left( {\cos \left( {{\phi _{\rm{D}}}} \right) + \cos \left( {{\phi _{\rm{A}}}} \right)} \right),
\end{equation}
and the rotation angle-determined $R_{2,k,i}$ is expressed by
\begin{equation}
		{R_{2,k,i}} = \frac{\pi }{2}{\left( {M - 2i + 1} \right)}{\left( {\cos {\left( {{\phi _{\rm{D}}}-{\theta _k}} \right)} + \cos {\left( {{\phi _{\rm{A}}}-{\theta _k}} \right)}} \right)}.
\end{equation}
\end{proposition}
\begin{IEEEproof}
	Applying Jensen's inequality to ${\mathbb E}\{S_{\rm BC}\}$ and following \cite{Y. Han-LIS}, we derive $S_{\rm BC,u}$.
\end{IEEEproof}

\begin{corollary}\label{Cor-1}
	When MRT is adopted for EC-RIS-assisted MISO systems, the upper bound of averaged SE, $ S_{\rm EC,u}$, can be derived from \eqref{Eq:SE-u} by setting $M=1$.
\end{corollary}

From \emph{Proposition \ref{Pro-1}} and \emph{Corollary \ref{Cor-1}}, the optimal reflection phase that maximizes the averaged SE is given by
\begin{equation}\label{Eq:gamma}
	{\gamma _k}=-\left( {k - 1} \right)M\pi \left( {\cos \left( {{\phi _{\rm{D}}}} \right) + \cos \left( {{\phi _{\rm{A}}}} \right)} \right),
\end{equation}
where $k\in\{1,2,\ldots,K\}$ for the BC-RIS and  $k\in\{1,2,\ldots,N_{\rm s}\}$ for the EC-RIS. This approach aims to coherently combine the received LoS signals that are reflected by the central elements of RIS blocks or by the independently controllable RIS elements. When optimal reflection phases are designed and channels $\bf g$ and $\bf G$ remain unchanged, the BC-RIS incurs an averaged SE loss, $\Delta {S_{\rm{u}}}={S_{{\rm{EC}},{\rm{u}}}} - {S_{{\rm{BC}},{\rm{u}}}}$, expressed by
\begin{equation}
		{  {\Delta {S_{\rm{u}}}	={\log _2}{\left( {\frac{{P\left( {{C_1}N_{{\rm s}}^2 + {C_2}} \right) + {\sigma ^2}}}{{P\left( {{C_1}{K^2}{{\left| {\sum\limits_{i = 1}^M {{e^{ - j{R_{2,k,i}}}}} } \right|}^2} + {C_2}} \right) + {\sigma ^2}}}} \right)}.}}		
\end{equation}
Because $| {\sum\nolimits_{i = 1}^M {{e^{ - j{R_{2,k,i}}}}} } |\le  {\sum\nolimits_{i = 1}^M |{{e^{ - j{R_{2,k,i}}}}}| } =M$, $\Delta {S_{\rm{u}}} \ge 0$ holds. To achieve the same averaged SE performance of the EC-RIS, i.e., $\Delta {S_{\rm{u}}} = 0$, we can finetune channels $\bf \bar{g}$ and $\bf \bar{G}$ by rotating RIS blocks to make ${\cos \left( {{\phi _{\rm{D}}}-{\theta _k}} \right) + \cos \left( {{\phi _{\rm{A}}}-{\theta _k}} \right)}=0$, yielding
\begin{equation}\label{Eq:theta_k}
	{\theta _k} =  {({\phi _{\rm{D}} + \phi _{\rm{A}}})}/{{\rm{2}}}-{\pi }/{2}.
\end{equation}
This rotation angle aligns the AoA and AoD specularly with respect to the direction orthogonal to the block. In the considered single-UE system, all blocks share the same rotation angle. However, when multiple UE are present, customizing specular reflection channel for each UE will result in different rotation angles for each block. Addressing this issue remains as our future works. In typical scenarios where ULAs of the BS and RIS are parallel and the coverage sector of the RIS is  $\frac{2\pi}{3}$, i.e., $\phi_{\rm{A}}=\frac{\pi}{2}$ and $\phi_{\rm{D}}\in [\frac{\pi}{6},\frac{5\pi}{6}]$, we have $|\theta_k|\in[0,\frac{\pi}{6}]$, indicating that a small rang of rotation angle is sufficient.

These analyses demonstrate that the rotatable BC-RIS can achieve the same averaged SE as the EC-RIS with a reduced number of reflection phase control circuits.

\subsection{Averaged Energy Efficiency Analysis}

While the BC-RIS and EC-RIS can achieve equivalent upper bounds of averaged SE, their PCs differ. To achieve a higher averaged EE with the BC-RIS, $P_{\rm BC} < P_{\rm EC}$ must hold. In the considered scenario, each RIS block rotates the same angle, denoted as $\theta _k = \theta$, as shown in \eqref{Eq:theta_k}. Thus, the PC of the BC-RIS can be simplified as
\begin{equation}\label{Eq:APP-B1}
	{P_{{\rm{BC}}}} = {P_0} + \xi P + K\left( {{P_1} + {P_2}} \right) + K\frac{{{M^2} - 1}}{4}\left| {{\theta }} \right|{P_{{\rm{unit}}}}.
\end{equation}
Given that this PC is determined by the size of the RIS block, we provide a segmentation scheme that minimizes $P_{\rm BC}$.
\begin{proposition}\label{Pro-2}
	When $N_{\rm s}$ is fixed and the block rotation angle is configured by \eqref{Eq:theta_k}, the number of blocks that minimize the PC of the BC-RIS can be given by
	{ {\begin{equation}\label{Eq:M-}
	{ {K^{\star} = \left\{ \begin{aligned}
			 \frac{N_{\rm s}}{2},\qquad\quad &\;     {\rm for}\;\frac{4}{5}P_{\rm ratio}\le \left| {{\theta}} \right| \le \frac{\pi}{6},\\
			 \sqrt {\frac{{N_{{\rm s}}^2\left| {{\theta }} \right|}}{{4{P_{{\rm{ratio}}}} - \left| {{\theta }} \right|}}}  ,&\;{\rm for}\;\frac{4P_{\rm ratio}}{{ {{N^2_{\rm{s}}} + 1} }}<\left| {{\theta }} \right| < \frac{4}{5}P_{\rm ratio},\\
			 1,\qquad\quad &\;{\rm for}\;0\le\left| {{\theta }} \right| \le \frac{4P_{\rm ratio}}{{ {{N^2_{\rm{s}}} + 1} }},
	\end{aligned}  \right.	}}
\end{equation}}}
where $P_{\rm ratio}=(P_1+P_2)/P_{\rm unit}$.
\end{proposition}
\begin{IEEEproof}
	Substituting $M = {N_{{\rm s}}}/K$ into \eqref{Eq:APP-B1} and calculate the derivative of ${P_{{\rm{BC}}}}$ with respect to $K$, we obtain \eqref{Eq:M-} via the monotonicity of ${P_{{\rm{BC}}}}$.
\end{IEEEproof}

In practice, $K^{\star}$ should be normalized to an integer when $4P_{\rm ratio}/( {{N^2_{\rm{s}} + 1} )}<\left| {{\theta}} \right| < \frac{4}{5}P_{\rm ratio}$ because it represents the number of RIS blocks. Nevertheless,  \emph{Proposition \ref{Pro-2}}  shows that the block segmentation strategy depends on the rotation angle. Intuitively, the PC saved by the reduced phase control circuit is linearly proportional to $M$. However, \eqref{Eq:APP-B1} reveals that the rotation angle-incurred power quadratically increases with $M$. Therefore, when $| {{\theta }} |$ grows, the block size should be smaller ($K$ should be larger), as shown in \eqref{Eq:M-}. In cases where the rotation angle is sufficiently small, i.e., $| {{\theta }} | \le 4P_{\rm ratio}/( {{N^2_{\rm{s}}} + 1} )$, the power consumed by the rotation angle and rotate control circuit is always less than that saved by the reduced phase control circuit. Thus, the block size can be maximized to $N_{\rm s}$.

Note that \emph{Proposition \ref{Pro-2}} is a general result for RIS block segmentation, it implies $P_{\rm unit}\ge \frac{24}{5\pi}(P_1+P_2)$ due to the fact that $|\theta|\le \pi/6$. In practical hardware implementations, this criterion can be simplified with different values of $P_{\rm unit}$. For instance, when the unit PC is sufficiently small, that is $P_{\rm unit}\le\frac{24}{\pi }( {{P_1} + {P_2}} )/( {N_{{\rm s}}^2 + 1} )$, $K_{\rm opt}=1$ always holds because ${4P_{\rm ratio}}/({{ {{N^2_{\rm{s}}} + 1} }})\ge \pi/6\ge |\theta|$.

In the general block segmentation where $P_{\rm unit}\ge \frac{24}{5\pi}(P_1+P_2)$, ${P_{{\rm{BC}}}}$ varies for different cases. To achieve a higher averaged EE, the BC-RIS should have a smaller PC than that of EC-RIS. Apart from $P_{\rm unit}$, the BC-RIS introduces another power coefficient $P_2$. We provide the following proposition to showcase the ranges of $P_2$ and $P_{\rm unit}$ that satisfy $E_{\rm BC}> E_{\rm EC}$.

\begin{proposition}\label{Pro-3}
	When the number of RIS elements is fixed, dividing the RIS into rotatable blocks can achieve a higher averaged EE when the PC of the rotate control circuit $P_2$ satisfies \eqref{Eq:Pro3} given in the next page.

\end{proposition}
\begin{IEEEproof}
	With a decreasing $P_{\rm unit}$, simpler RIS block segmentation schemes and their maximum $P_{\rm BC}$ with respect to $|\theta|$ can be derived. By letting the maximum $P_{\rm BC}$ smaller than $P_{\rm EC}$, this proof can be completed.
\end{IEEEproof}
\begin{figure*}
	\begin{subequations}\label{Eq:Pro3}
		{{\begin{numcases}{}
			{P_2} < {P_1} - {{{\pi }}{P_{{\rm{unit}}}}}/{8}, \;\quad\qquad\qquad\qquad\qquad~  {\rm for}\; {P_{{\rm{unit}}}} \ge {24}({P_1} + {P_2})/({{5\pi }}), \label{Eq:Pro3-1} \\
			{P_2} <{\left( {{P_{{\rm{unit}}}} - {{12{P_1}}}/{\pi }} \right)^2}{\pi }/({{24{P_{{\rm{unit}}}}}}), \qquad\quad{\rm for}\;{{24}}({P_1} + {P_2})/{{ \left( \pi{N_{{\rm s}}^2 + \pi} \right)}} < {P_{{\rm{unit}}}} < {{24}}({P_1} + {P_2})/({{5\pi }}),\label{Eq:Pro3-2} \\
			{P_2} <\left( {{N_{{\rm s}}} - 1} \right){P_1} - {{\left( {N_{{\rm s}}^2 - 1} \right)\pi }{P_{{\rm{unit}}}}}/{{24}},   \;\;\; \; {\rm for}\;0 \le {P_{{\rm{unit}}}} \le {{24}}({P_1} + {P_2})/{{ \left( \pi{N_{{\rm s}}^2 + \pi} \right)}}. \label{Eq:Pro3-3}
		\end{numcases}}}
	\end{subequations}
	\hrulefill
	\vspace{-0.5cm}
\end{figure*}

\emph{Proposition \ref{Pro-3}} offers guidance for the hardware implementation of the rotatable BC-RIS, particularly in terms of PC. Given a preset $P_1$, selecting appropriate values for $P_2$ and $P_{\rm unit}$ that satisfy the inequalities shown in \eqref{Eq:Pro3} will enable achieving higher EE.

\section{Numerical Results}\label{sec:4}
In this section,  numerical results are presented to evaluate the averaged SE/EE of the EC- and BC-RIS.  The BS is equipped with $N_{\rm b}=32$ antennas. For the power-related parameters, we set ${\xi =1.2}$,  ${P_0=12}$ W, ${P_1=0.12}$ W. With the noise power normalized as $\sigma^2=1$, the transmit power can be calculated as $P=10^{{\sf snr} /10}$, where ${\sf snr}$ is the signal-to-noise ratio (SNR).

\begin{figure}
	\centering
	\includegraphics[width=0.43\textwidth]{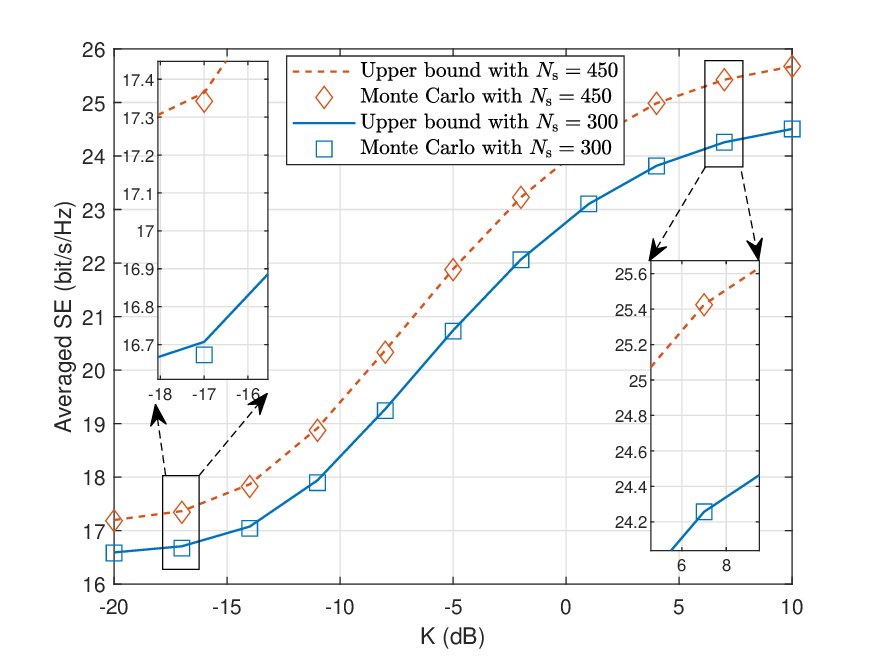}
	\caption{Tightness of the upper bound of averaged SE with the rotatable BC-RIS.}
	\label{Fig.SE-K}
	\vspace{-0.5cm}
\end{figure}

The upper bound and Motel Carlo result of the averaged SE with the rotatable BC-RIS are presented in Fig. \ref{Fig.SE-K}. The small performance gap reveals that our analytical result is a tight bound. Notably, when the Rician factor or the number of RIS elements increases, this gap is further narrowed. 

\begin{figure}
	\centering
	\subfigure[SE]{
		\label{Fig.SE} 
		\includegraphics[width=0.233\textwidth]{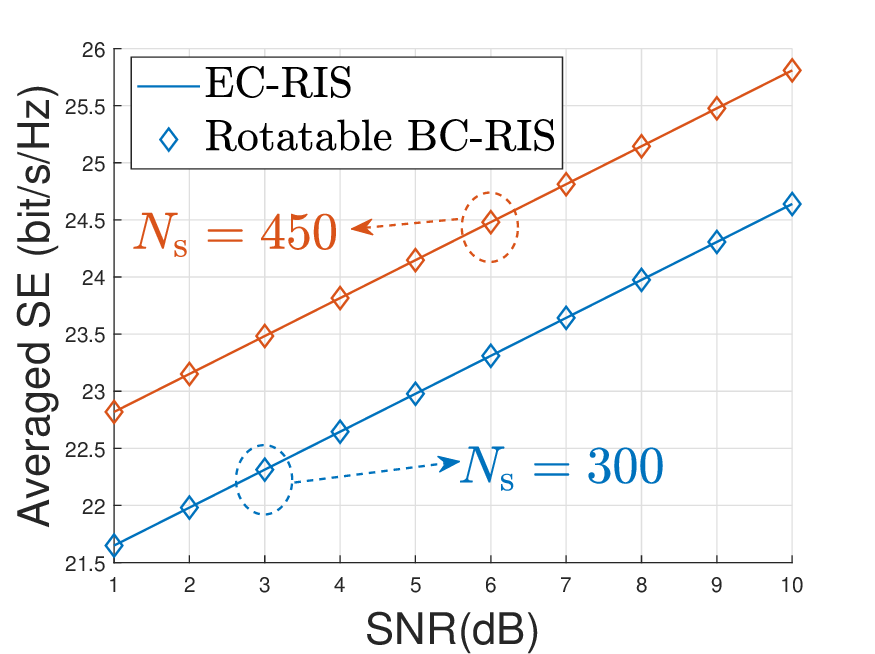}}
	\subfigure[EE]{
		\label{Fig.EE-SNR}
		\includegraphics[width=0.233\textwidth]{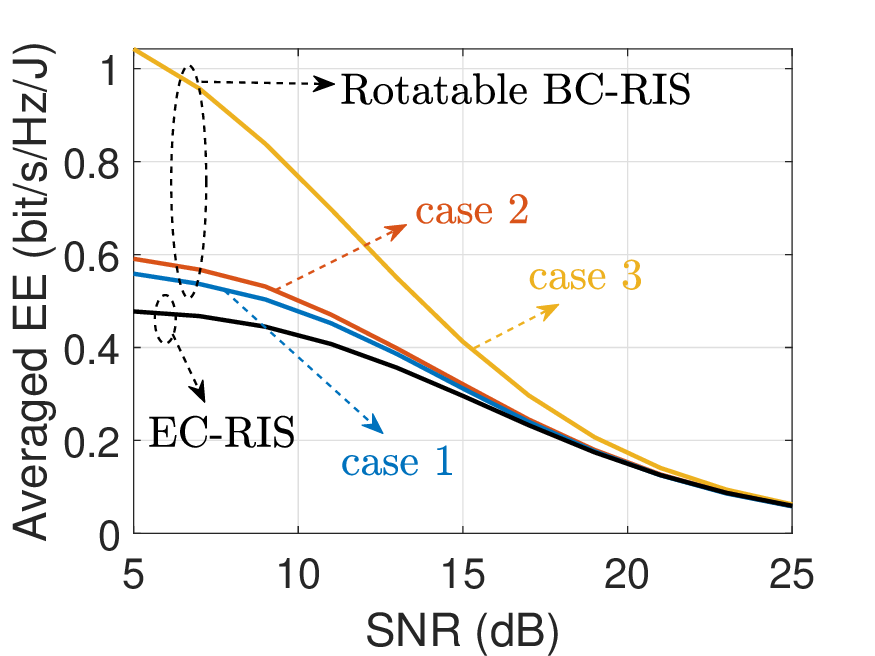}}
	\caption{Averaged SE and EE comparison for the EC-RIS and the rotatable BC-RIS.}
	\label{Fig.ratio} 
	\vspace{-0.5cm}
\end{figure}

The average SE of the EC-RIS and the rotatable BC-RIS are shown in Fig. \ref{Fig.SE} to evaluate the tightness of these schemes.  Although the \emph{conventional} BC-RIS results in a SE loss due to the loss of degree of freedom for reflection phase design  \cite{BC-RIS}, it is shown that the \emph{rotatable} BC-RIS achieves performance closely matching that of the EC-RIS. These results indicate that the averaged SE loss caused by reduced phase shifter controller can be well compensated by increasing an rotation mechanism to customize a specular reflection channel.

The advantages of the BC-RIS are presents in Fig. \ref{Fig.EE-SNR}, where $\{{P_2 = 0.108 \,{\rm W}}, {P_{\rm unit}=0.821 \,{\rm W}}\}$ in case 1, $\{{P_2 = 0.215\,{\rm W}}, \ {P_{\rm unit}=0.548\,{\rm W}}\}$ in case 2, and $\{P_2 = 0.430\,{\rm W}, {P_{\rm unit}=0.003\,{\rm W}}\}$ in case 3 are designed according to \eqref{Eq:Pro3-1}, \eqref{Eq:Pro3-2}, and \eqref{Eq:Pro3-3}, respectively. In the low SNR regime, the BC-RIS can achieve significantly higher EE than that of the EC-RIS, especially in case 3, where $P_{\rm unit}$ is sufficiently small. These results are intuitive because the BC-RIS fabricated with lighter materials has smaller rotation PC. Considering the lighter weight and smaller size of the RIS at higher frequency bands, our proposal is highly attractive for millimeter wave and terahertz communications. In the high SNR regime where the transmit power dominates the system PC, our proposal achieves the same EE as the EC-RIS because the PC saved by the rotatable BC-RIS accounts for a small proportion of the systems PC.


\section{Conclusion}\label{sec:5}
In this letter, we introduced the concept of a BC-RIS featuring mechanically rotatable blocks to emulate the performance of an EC-RIS. Specifically, we demonstrated that by adjusting the rotation angle of each RIS block to customize a specular reflection channel, the BC-RIS can achieve the same upper bound of averaged SE as the EC-RIS, but with fewer phase shifter controllers. Given a PC model for the BC-RIS, we derived the optimal segmentation of RIS blocks that minimizes the power required to achieve the maximum averaged SE. To ensure that the BC-RIS achieves higher averaged EE compared to the EC-RIS, we provided the ranges of rotation-related PC to guide the hardware design of the BC-RIS. Simulation results demonstrated that the BC-RIS can achieve the same averaged SE as the EC-RIS. Moreover, by carefully setting PC-related parameters, the BC-RIS exhibited higher EE compared to the EC-RIS.

\begin{small}

\end{small}
%


\begin{thebibliography}{30}

\bibitem{6G}C. -X. Wang \emph{et al.}, ``On the road to 6G: Visions, requirements, key technologies, and testbeds,'' \emph{IEEE Commun. Surveys. Tuts.}, vol. 25, no. 2, pp. 905-974, 2nd Quar., 2023.

\bibitem{E. Basar}E. Basar \emph{et al.}, ``Reconfigurable intelligent surfaces for 6G: Emerging hardware architectures, applications, and open challenges,'' \emph{IEEE Vehicular Tech. Mag.}, early access, 2024.

\bibitem{App_EE}S. Zhang, B. Di, and H. Zhang, ``Energy efficient symbol level precoding design in a RIS assisted communication system," \emph{IEEE Commun. Lett.}, vol. 28, no. 5, pp. 1141--1145, May 2024.


\bibitem{Q. Wang-IRS}Q. Wang, Q. Chen, Y. Zhang, and H. Wang, ``Learning-based intelligent reflecting surface-aided cell-free massive MIMO systems," \emph{IEEE Trans. Vehicular Tech.}, vol. 72, no. 9, pp. 12338--12342, Sep. 2023.

\bibitem{app-22-ZLin}Z. Lin \emph{et al.}, ``Refracting RIS aided hybrid satellite-terrestrial relay networks: Joint beamforming design and optimization,” \emph{IEEE Trans. Aerosp. Electron. Syst.}, vol. 58, no. 4, pp. 3717-3724, Aug. 2022.

\bibitem{TVT-2}D. K. P. Asiedu and J. -H. Yun, ``Multiuser NOMA with multiple reconfigurable intelligent surfaces for backscatter communication in a symbiotic cognitive radio network,'' \emph{IEEE Trans. Vehicular Tech.,} vol. 72, no. 4, pp. 5300-5316, Apr. 2023.

\bibitem{Y. Han-LIS}Y. Han \emph{et al.}, ``Large intelligent surface-assisted wireless communication exploiting statistical CSI,'' \emph{IEEE Trans. Vehicular Tech.,} vol. 68, no. 8, pp. 8238-8242, Aug. 2019.

\bibitem{CC-1}W. Chen, C.-K. Wen, X. Li, and S. Jin, ``Channel customization for joint Tx-RISs-Rx design in hybrid mmWave systems,’’ \emph{IEEE Trans. Wireless Commun.}, vol. 22, no. 11, pp. 8304-8319, Nov. 2023.


\bibitem{FA-kit}K. K. Wong \emph{et al.} ``Fluid antenna systems,'' \emph{IEEE Trans.
Wireless Commun.}, vol. 20, no. 3, pp. 1950–1962, Mar. 2021.

\bibitem{MA}L. Zhu, W. Ma, and R. Zhang, ``Modeling and performance analysis for movable antenna enabled wireless communications,'' \emph{IEEE Trans. Wireless Commun.}, vol. 23, no. 6, pp. 6234--6250, Jun. 2024.

\bibitem{FP-MIMO}J. Zheng \emph{et al.} ``Flexible-position MIMO for wireless communications: Fundamentals, challenges, and future directions,''  \emph{IEEE Wireless Commun.}, early access, 2024.

\bibitem{B. Zheng}B. Zheng and R. Zhang, ``Intelligent reflecting surface-enhanced OFDM: Channel estimation and reflection optimization," \emph{IEEE Wireless Commun. Lett.}, vol. 9, no. 4, pp. 518--522, Apr. 2020.

\bibitem{BC-RIS}X. Yang \emph{et al.}, ``Performance evaluation for subarray-based reconfigurable intelligent surface-aided wireless communication systems,’’ in \emph{Proc. IEEE Global Communications Conference}, Dec. 2023, pp. 1-6.



\bibitem{J.Wang}J. Wang \emph{et al.}, ``Reconfigurable intelligent surface: Power consumption modeling and practical measurement validation,''  \emph{IEEE Trans. Commun.}, early access, Mar. 2024.





\end{thebibliography}
\end{document}